\documentclass[preprint,12pt]{elsarticle}

\usepackage{siunitx}
\usepackage{amsmath}
\usepackage{amssymb}
\usepackage{amsfonts}

\usepackage{amssymb}

\newcommand*{\anu}{\ensuremath{\bar{\nu}_e}}
\newcommand*{\anumu}{\ensuremath{\bar{\nu}_\mu}}

\newcommand{\boss}[2]{\ensuremath{\rlap{\kern-2.5pt\ensuremath{\overset{\scriptscriptstyle(-)}{\phantom{#1}}}}{\ensuremath{{#1}_{#2}}}}}

\journal{Nuclear Physics B}

\begin{document}

\begin{frontmatter}



\title{Light Sterile Neutrinos in Particle Physics: Experimental Status}

\author[label1,label2]{Thierry Lasserre}
\address[label1]{Commissariat a l'\'energie atomique et aux \'energies alternatives, Centre de Saclay, IRFU, 91191 Gif-sur-Yvette, France}
\address[label2]{Astroparticules et Cosmologie APC, 10 rue Alice Domon et L\'eonie Duquet, 75205 Paris cedex 13, France}

\author{}

\address{}

\begin{abstract}
Most of the neutrino oscillation results can be explained by the three-neutrino paradigm. 
However several anomalies in  short baseline oscillation data, corresponding to an L/E of about 1~m/MeV,  
could be interpreted by invoking a hypothetical fourth neutrino. This new state would be 
separated from the three standard neutrinos by a squared mass difference $\Delta m_{new}^2  \sim 0.1- 1\,\si{eV^2}$ 
and would have mixing angles of $\sin^2 2\theta_{ee} \gtrsim 0.01$ and $\sin^2 2\theta_{\mu e} \gtrsim 0.001$, in the electron disapperance and appearance channels, respectively. 
 This new neutrino, often called sterile, would not feel 
standard model interactions but mix with the others. Such a scenario calling for new physics 
beyond the standard 
model has to be either ruled out or confirmed with new data. After a brief review of the 
anomalous oscillation results 
we discuss the forthcoming laboratory experiments aiming to clarify the situation.

\end{abstract}


\end{frontmatter}


\section{Introduction}
\label{sec:intro}

The well established standard neutrino oscillation framework satisfactory
explains most of neutrino data. It relies on three flavours ($\nu_e, \nu_\mu, \nu_\tau$),
mixture of three mass states ($\nu_1, \nu_2, \nu_3$) separated by squared mass differences
of $\Delta m_{21}^2=\Delta m_{\mathrm{sol}}^2=7.50_{-0.20}^{+0.19} \times 10^{-5}\,\si{eV^2}$
and $\mid\Delta m_{31}^{2}\mid\approx\mid\Delta m_{32}^{2}\mid=\Delta m_{\mathrm{atm}}^2=2.32_{-0.08}^{+0.12} 
\times 10^{-3}\,\si{eV^2}$~\cite{Beringer:1900zz},
where "sol'' and "atm'' stand historically for solar and atmospheric experiments
providing compelling evidence for neutrino oscillation (see~\cite{Bilenky:2014} and references therein for a recent review).
Beyond this minimal extension of the standard model, anomalous results have been reported
in LSND~\cite{Aguilar:2001ty}, MiniBooNE~\cite{AguilarArevalo:2008rc,AguilarArevalo:2010wv},
and radioactive source experiments~\cite{GallexSage, Giunti:2006bj,Giunti:2010zu,Giunti:2012tn}.
In addition a new evaluation of the reactor neutrino fluxes~\cite{Mueller2013,Huber2011} led to a reinterpretation 
of the results of short baseline reactor experiments~\cite{Mention:2011}, the so-called Reactor Antineutrino Anomaly.

If not related to non understood experimental issues, results of the global fit
of short-baseline neutrino oscillation experiments (see~\cite{Kopp:2011} for instance) show that 
the data can be explained by the addition of one or two sterile neutrinos to the three active neutrinos 
of the standard model, the so-called (3+1) and (3+2) scenarios, respectively. However some tension remains 
between appearance and disappearance data in the global fits, see~\cite{Kopp:2013}. 

It is worth noting that sterile neutrinos would affect the oscillation probabilities of the active flavors 
and therefore could influence cosmological processes~\cite{CosmoSterileNu}. These aspects
 won't be further discussed in this experimental review focusing on terrestrial 
experiments, but more details can be found in~\cite{CosmoReview} .

\section{Anomalous Oscillation Results and Sterile Neutrinos} 
\label{sec:anomalies}

In this section we focus on neutrino oscillation results with an L/E of about 1~m/MeV. A comprehensive
review of all short baseline oscillation results and detailed statments on the current oscillation anomalies 
 can be found in~\cite{WhitePaper2012}.

In 1995 the LSND experiment reported an excess in the $\bar{\nu}_{\mu}\rightarrow\bar{\nu}_e$ appearance channel~\cite{Aguilar:2001ty}. 
A similar experiment, KARMEN~\cite{Armbruster:2002mp}, did not report such an excess, however.
In 2002 the MiniBooNE experiment confirmed this excess in both $\nu_e$ to $\nu_\mu$ and 
$\anu$ to $\anumu$ channels~\cite{AguilarArevalo:2008rc,AguilarArevalo:2010wv}. 
The MiniBooNE results will be soon complemented by using a 170-ton LAr TPC in the same neutrino beam; 
the MicroBooNE experiment~\cite{Ignarra2011} will check if the low-energy excess is due to $\nu_e$ charged current 
quasielastic events. 
Event rates measured by many reactor experiments at short distances, when compared with a newly evaluated antineutrino 
flux, are indicating the disappearance of $\anu$~\cite{Mention:2011}. In addition the results from the gallium solar neutrino 
calibration experiments reported also a deficit of $\nu_e$ in a similar L/E range~\cite{Giunti:2006bj,Giunti:2010zu,Giunti:2012tn}. 

The individual significances of these anomalies lie between 2.5 to 3.8 $\sigma$, and these results, not fitting 
 the three-neutrino-flavor framework, are difficult to explain by systematics effects. 
If not experimental artifacts it is puzzling 
that each of them could be explained by oscillation to sterile neutrinos with a large mass squared difference, 
$\Delta m_{new}^2 \gtrsim 0.1 \,\si{eV^2}$, corresponding to an L/E of about 1~m/MeV. 

Indeed the minimal neutrino mixing scheme provides only two squared-mass differences. A third one 
 would be required for new short-baseline neutrino oscillations. It then require the introduction of a sterile 
neutrino $\nu_{s}$~\cite{Bilenky:1998dt,Maltoni:2004ei,Strumia:2006db,GonzalezGarcia:2007ib}.
The minimal model consists of a hierarchical 3+1 neutrino mixing, acting as a perturbation of the standard three-neutrino 
mixing in which the three active neutrinos $\nu_{e}$, $\nu_{\mu}$, $\nu_{\tau}$ are mainly composed of three massive 
neutrinos $\nu_1$, $\nu_2$, $\nu_3$ with light masses $m_1$, $m_2$, $m_3$. The sterile neutrino would mainly be 
composed of a heavy neutrino $\nu_{4}$ with mass $m_{4}$ such that $\Delta{m}^2_{\text{new}} = \Delta{m}^2_{41}$, 
and $m_{1} \,,\, m_{2} \,,\, m_{3} \ll m_{4}$.

In 3+1 neutrino mixing, the effective flavor transition and survival probabilities in short-baseline neutrino oscillation 
experiments are given by
\begin{equation}
P_{\boss{\nu}{\alpha}\to\boss{\nu}{\beta}}^{\text{new}}
=
\sin^{2} 2\theta_{\alpha\beta}
\Delta_{41}
\,
(\alpha\neq\beta)
\,,
P_{\boss{\nu}{\alpha}\to\boss{\nu}{\alpha}}^{\text{new}}
=
1
-
\sin^{2} 2\theta_{\alpha\alpha}
\Delta_{41}
\end{equation}
where 
$\Delta_{41}=\sin^{2}\left( \frac{\Delta{m}^2_{41} L}{4E} \right)$, 
and for $\alpha,\beta=e,\mu,\tau,s$,
with the transition amplitudes
\begin{equation}
\sin^{2} 2\theta_{\alpha\beta}
=
4 |U_{\alpha4}|^2 |U_{\beta4}|^2
\,,
\\
\sin^{2} 2\theta_{\alpha\alpha}
=
4 |U_{\alpha4}|^2 \left( 1 - |U_{\alpha4}|^2 \right)
\,.
\end{equation}

The interpretation of both LSND and MiniBooNE anomalies in terms of light sterile neutrino 
oscillations requires 
mixing of the sterile neutrino with both electron and muon neutrinos. 
In addition, both OPERA and ICARUS experiments recently reported negative results for the 
search $\nu_e$ from 
the  $\nu_\mu$  CNGS beam~\cite{Agafonova:2013, Segreto:2013},  although not testing 
fully the relevant space of 
oscillation parameters. Therefore when considering all data together 
no satisfactory global fit can be obtained (see ~\cite{Kopp:2013} for instance). 
This is mainly due to the non-observation of 
$\nu_\mu$ disappearance 
at the eV-scale~\cite{Dydak:1984}, that is a generic prediction if the LSND signal implies a 
sterile neutrino. This
negative results is not strong enough to rule out this hypothesis, however. 

All these facts motivate the experimental program being briefly summarized in this review. 
In what follows, we focus 
on the  3 active +1 sterile neutrino mixing scheme with  $\Delta m^2_\mathrm{new}$ of the order 
of \SIrange{0.1}{1}{eV^2}.

\section{Clarification of the Anomalies: Experimental Program}
\label{sec:projects}

To definitively test the short baseline oscillation hypothesis the new experiments must be sensitive 
to an oscillation pattern either in the energy spectrum,
or in the spatial distribution of the neutrino interactions, or both.
To cover the $\Delta m^2$ region of \SIrange{0.1}{1}{eV^2} with MeV/GeV neutrinos the distance 
between the emitter and the detector has to be on the scale of~1-10~m~/~1-10 km, respectively. 
Statistical and systematics uncertainties must be at the level of a few percents or less.
Such an experiment could be performed close to nuclear reactors, with intense 
radioactive sources used as neutrino emitters, or with accelerator based experiments. We review below
the various projects that have been proposed to clarify the neutrino anomalies, leaving out R\&D efforts.

\subsection{Reactor-based Proposals}
\label{sec:reactor}

Nuclear reactors are very intense sources of 1-10 MeV electron antineutrinos. In the 80's their expected fluxes were 
obtained with a precision of 5\% through the measurement of the integral $\beta$-spectra of uranium and 
plutonium isotopes irradiated into a reactor core, followed by their phenomenological conversion into 
$\anu$ spectra~\cite{SchreckU5Pu9,SchreckPu9Pu1}. But in 2011 this prediction of was corrected leading to 
an increase of  the emitted flux by about 4\%, with a similar precision~\cite{Mueller2013,Huber2011}. 
The revised comparison of the latest with the measured rate 
of interactions in detectors located at 100 m or less from the cores revealed the Reactor Antineutrino 
Anomaly~\cite{Mention:2011}. 
It is worth noting that there remains some lack of knowledge of the reactor neutrino fluxes. It has been recently 
pointed out that the detailed treatment of forbidden transitions in the computation of reactor neutrino spectra 
may lead to an increase of the systematic uncertainty by a few percents~\cite{Hayes:2014}. 
Moreover, while writing this article a new deviation with respect to the expected reactor neutrino spectral shape predictions~\cite{SchreckU5Pu9,SchreckPu9Pu1,Mueller2013,Huber2011}  has been announced
by the RENO and Double Chooz collaborations at the Neutrino 2014 conference~\cite{RENO5MeV:2014,DC5MeV:Nu2014,DC5MeV:2014}, 
and confirmed later by the Daya Bay collaboration at the ICHEP 2014 conference~\cite{DB5MeV:2014}. 
This deviation in the prompt signal energy spectrum is being observed between about 4 to 7 MeV (visible energy) 
with a significance  of more than 3 standard deviations. The origin of this structure is still unknown. 
Therefore further investigations of reactor neutrino spectra as well as more precise data are needed.

New reactor experiments searching for short baseline oscillation, with 
L/E$\sim$1 m/MeV, should first look for an oscillation pattern imprinted in the energy distribution of
 events. Of course the analysis must be complemented by an integral rate measurement. 
According to global fits the relevant range of oscillation lengths, $L_{osc} \sim ￼2.5 E/\Delta m_{new}^2$ is 
between 1 and 10 meters. Therefore short baselines, a few ten's of meters, are mandatory and compact 
reactor cores, with typical dimensions of 1 m, are preferable in comparison with larger commercial reactors 
to prevent  washing out the hypothetical oscillation pattern at the L/E's of interest.
\begin{table}[!h]
\begin{center}
\begin{tabular}{lr|c|ccc}
\hline
Projects & Ref  & $P_{th}$ & $M_{target}$ & $L$ & Depth \\
 & & (MW) & (tons) & (m) & (m.w.e.) \\ \hline\hline
Nucifer & \cite{Nucifer} & 70 & 0.75 & 7 & 13 \\ 
Stereo & \cite{WhitePaper2012} &  50 & 1.75 & [8.8-11.2] & 18 \\ 
Neutrino 4 & \cite{Neutrino4}&  100 & 2.2 & [6-12] & few \\ 
DANSS & \cite{DANSS} &  3 & 0.9 & [9.7-12.2] & 50 \\
Solid & \cite{Solid} &  [45-80] & 3 & [6-8] & 10 \\
Hanaro &  &  30 & 0.5 & 6 & few \\
US project & \cite{USproj} &  20-120 & 1 \& 10 & 4 \& 18 &  few \\
CARR & \cite{CARR} &  60 & - & 7 \& 15 & few \\
\hline
\end{tabular}
\caption{Main features of proposed reactor experiments.}
\label{tab:reactorexp}
\end{center}
\end{table}
Experimentally the detection technique of most experiments in preparation relies on the inverse $\beta$-decay (IBD) reaction, 
$\anu$ + p $\rightarrow$ e$^+$ + n, where the positron carries out the $\anu$ energy while the 
neutron tagging provides a discriminant signature against backgrounds.
Indeed an accidental pair from $\gamma$-ray radioactivity contaminants or induced by the reactor core, followed 
by a neutron capture or a high energy $\gamma$ from the core could easily mimic the signal. This background 
can partially be suppressed through passive shielding while the remaining contribution can be measured in-situ 
at the analysis stage, leading to an increase of the uncertainty due to statistical fluctuations of the background rate, however.
Correlated backgrounds induced by cosmic rays can also alter the signal. By definition a single correlated 
event can mimic the IBD process. All currents projects are foreseen at shallow depths or even at the surface, 
the latter case being extremely challenging and not yet experimentally demonstrated at the desired precision.
More problematic could be the possible correlated backgrounds induced by the reactor core itself. It must be
suppressed through passive shielding, depending strongly on the site configuration and on the type of reactor
core. This background superimposes on the top of the signal and it cannot be measured in situ, unfortunately. 
It is therefore mandatory to optimize the experimental setup through simulation to minimize it, while taking 
large safety margins due to the difficulty of assessing the remaining contribution in the fiducial volume.
Table~\ref{tab:reactorexp} provides a list of current projects being carried out at reactors. The Nucifer 
experiment~\cite{Nucifer}  is currently taking data close to the Osiris nuclear reactor in Saclay. Though not optimized 
for a sterile neutrino search it could provide first new constraints by 2015. The Stereo experiment~\cite{WhitePaper2012} 
 will be constructed next to the ILL reactor in Grenoble in 2014 and aims taking data middle of 2015.
The DANSS~\cite{DANSS}  and Neutrino4~\cite{Neutrino4} experiments are under construction in Russia and should 
provide first data in 2015.  Finally a comprehensive project for searching sterile neutrinos at reactor 
in US is currently in its R\&D phase~\cite{USproj}; depending on its approval schedule it could provide first results by 2016.
All these experiments are designed to test the space of parameters deduced from the interpretation of 
reactor antineutrino anomaly through the existence of light sterile neutrinos.

\begin{figure}[h!]
\centering 
\includegraphics[width=1\linewidth]{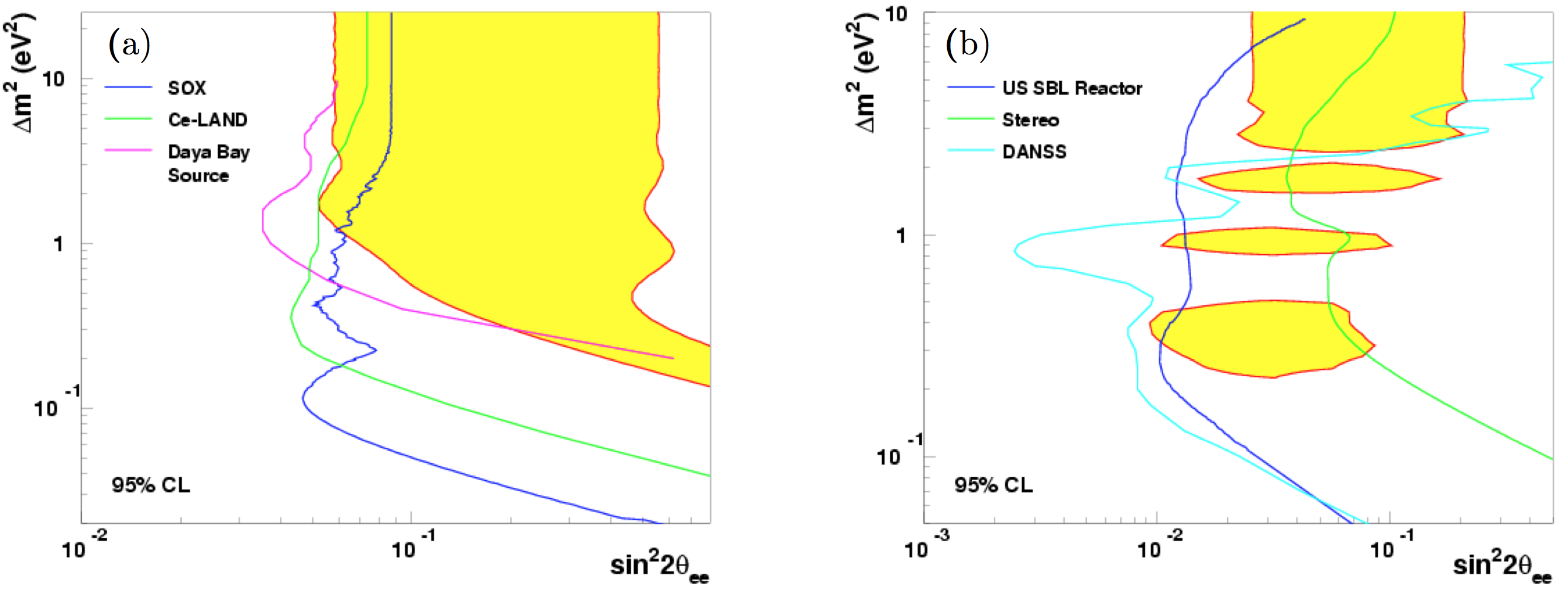}
\caption{\label{fig:1} Projected sensitivity curves for proposed neutrino generator and reactor 
experiments plotted against the global fits for the gallium anomaly and reactor anomaly respectively, 
based on the compilation of~\cite{SnowMass2013}.}
\end{figure}

\subsection{Neutrino Generator Proposals}
\label{sec:nugen}

In the experiments performed to calibrate the radiochemical solar neutrino detectors SAGE and GALLEX 
 the number of measured capture events for neutrinos from artificial sources of $^{51}$Cr and $^{37}$Ar are 
below  the expectations, the average ratio of the measured-to-expected capture-rate being 0.87 $\pm$ 0.05 [5]. 
New experiments have been proposed to clarify this anomaly, using a very intense $^{51}$Cr neutrino generator 
at Baksan (3 MCi) or next to the Borexino detector (10 MCi). 
On the other hand an experiment using 100 kCi of $^{144}$Ce-$^{144}$Pr could be used next to a large
liquid scintillator detector, such as Borexino or KamLAND, to directly test the reactor antineutrino anomaly. 

Those projects aims to search for an energy-dependent oscillating pattern in event spatial distribution of 
active neutrino interactions that would 
unambiguously determine neutrino mass differences and mixing angles if oscillation to light sterile neutrinos 
is the explanation of the gallium and/or reactor neutrino anomalies. We review below these proposals.

\subsubsection{$^{51}$Cr-based Neutrino Generator}

A neutrino source uses the electron capture process to produce monoenergetic neutrinos. 
Several neutrino sources have already been produced to calibrate radiochemical solar neutrino
experiments. Two nuclei are usually considered: $^{51}$Cr and  $^{37}$Ar.
The  $^{51}$Cr decays with a \SI{27.7}{day} half-life, producing mainly \SI{753}{keV} neutrinos,
and in \SI{10}{\%} of decays \SI{433}{keV} neutrinos with a \SI{320}{keV} gamma,
while the $^{37}$Ar produces \SI{814}{keV} neutrinos in any case with a \SI{35}{day} half-life.
The $^{37}$Ar is therefore more suitable from the point of vue of heat and shielding issues,
and benefits also of slightly longer half-life and slightly higher energy. Still chromium is much easier to handle.
Both isotopes have to be produced by neutron irradiation in a nuclear reactor,
through $^{50}$Cr (n,$\gamma$) $^{51}$Cr process and $^{40}$Na (n,$\alpha$) $^{37}$Ar process respectively.
Moreover, the (n,$\alpha$) reaction has a threshold requiring irradiation with fast neutron.

The main drawback of neutrino source relies in the detection process, elastic scattering off electrons.
The cross section of this process is low and the detection is very sensitive to backgrounds.
Currently only Borexino, design to study solar neutrinos, has shown a low enough background control.
The unique extreme radiopurity achieved in the liquid scintillator medium  allows to control the irreducible contribution 
of $^7$Be solar neutrinos. The experiment will consist in counting the number of observed events at each 
detector location and to compare it to the expectation without oscillations. The position of 
each event can be reconstructed with a precision of $\sim$12~cm at 1~MeV, which is enough 
for the range of $\Delta m^2$ of interest and smaller than the size of the source, a few tens of centimeters. 
The SOX experiment~\cite{SOX} will perform such an measurement with 
a \SI{10}{MCi} $^{51}$C source irradiated either in Russia (PA Mayak) or in US, and deployed at 8.25 m 
from the center of the Borexino detector in 2016/17.

At Baksan another technique is being pursued. Based on the technology developed for the SAGE solar neutrinos
experiment a $^{51}$Cr source could be placed at the center of a target, containing 50 t of liquid metallic gallium 
divided in two areas, an inner 8-ton zone and an outer 42-ton zone. 
The ratio of the two measured capture rate to  its expectation could
sign an oscillation, although not as precisely as for the oscillometry performed in a liquid scintillator detector. 
This is a well-proven technique free of backgrounds, however. Furthermore it would necessitate a lower 
activity, 3 MCi, more easy to achieve from standard irradiation in research reactors and logistic issues 
would be easier to organize since both the source and the detector would remain in Russia.

\subsubsection{$^{144}$Ce-$^{144}$Pr-based Antineutrino Generator}

An antineutrino source uses the $\beta^-$ decay process to produce
a non monoenergetic neutrino spectrum. Antineutrinos allow the use of Inverse Beta Decay (IBD)
as detection process: $\anu{} + p \rightarrow e^+ + n$.
At a few MeV's it has the advantage of a higher cross-section with respect to neutrino scattering 
off electrons, by roughly one order of magnitude. Furthermore the time and space coincidence between 
positron and neutron allow a very effective tagging of the process, leading to much easier background rejection.

The main drawback is the \SI{1.8}{MeV} energy threshold requiring a high Q-value $\beta^-$ 
decay. Since the period and the Q-value are strongly anticorrelated for $\beta^-$ decay,
this requirement leads to nuclei with a period shorter than the day,
preventing the effective production and use of an antineutrino source based on a single isotope.
The solution relies on the use of a cascade of two $\beta^-$ decays,
the father having a long period (month or year scale) and the daughter having a Q-value 
above the IBD threshold, as high as possible to maximize the event rate. 
Several pairs of isotope have been identified but we'll focus on the best option. 

The CeLAND and CeSOX experiments plan to use 100 kCi of $^{144}$Ce in 
KamLAND~\cite{PBqSource, CeLAND} and Borexino~\cite{PBqSource, CeSOX}. 
Cerium was chosen because of its high $Q_\beta$, its $\sim$4\% abundance in fission products 
of  uranium and plutonium, and finally for engineering considerations related 
to its possible extraction of rare earth from regular spent nuclear fuel reprocessing 
followed by a customized column chromatography. 
While not minimizing the difficulty of doing this, the nuclear industry does have the
technology to produce sources of the appropriate intensity, at a high purity level.
The goal is to deploy the $^{144}$Ce radioisotope about 10 m away from the detector center and to search 
for an oscillating  pattern in both event spatial and energy distributions that would determine 
neutrino mass differences and mixing angles through an unambiguously.
Thanks to available pressing technics the source fits inside a $<$15~cm-scale capsule,  
small enough to consider the Cerium volume as a point-like source. 
For comparison the vertex reconstruction is $<$15 cm.  
$^{144}$Ce has a low production rate of high-energy $\gamma$ rays ($>$ 1 MeV) 
from which the $\anu$ detector must be shielded to limit background events.  
Backgrounds are of two types, those induced by the environment or detector, 
and those due to the source (attenuated by a 20 cm tungsten shielding). Eventually
backgrounds are expected to be negligible thanks to the strong IBD signature. 

The logistic for transporting the source from the production site, PA Mayak in Russia, to the detector 
site is a major issue for such an experiment due to the necessary time required to certify the transport 
containers. This is a drawback for deploying quickly a  100 kCi  $^{144}$Ce source in KamLAND. Since
transportation to Italy is easier the CeSOX experiment could take data as early as end of 2015.

\subsubsection{Tritium-based Experiments}
A  new neutrino $\nu_{4}$ heavier than the three active neutrinos should leave an 
imprint in the $\beta$-spectrum of experiments measuring the absolute masses of active neutrinos, 
such as the forthcoming KATRIN experiment~\cite{katrin}. The detectors aims as measuring
precisely the high energy tail of the tritium $\beta$-decay spectrum by combining an intense molecular 
tritium source with an integrating high-resolution spectrometer. 
The projected sensitivity of the experiment on the effective electron neutrino mass is 200 meV at 90\% C.L. 
The detection principle is to search for a distortion at the high energy endpoint of the electron spectrum of 
tritium $\beta$-decay, since its shape is a priori very precisely understood. Any shape distortion due to 
decays involving an heavier neutrino could sign the existence of a sterile neutrino state.
As designed the KATRIN experiment can probe part of the current allowed region of the reactor antineutrino anomaly, 
especially for $\Delta m_{new}^2  > 1\,\si{eV^2}$ , with 3 years of data-taking~\cite{Formaggio2011, Esmaili2012}.
First results are expected in 2016.

\subsection{New Accelerator-based Proposals}
\label{sec:intro}

Over the last years a large experimental program is being prepared to search for sterile neutrinos using neutrino beams at 
CERN or Fermilab, or the spallation neutron source at Oak Ridge. We briefly review the various projects, sorting them by 
the processes creating the neutrinos.

\subsubsection{Isotope Decay at Rest}

A huge statistics of $\anu$ from the $\beta$-decay of $^{8}$He could be obtained through the development of a high-power 
cyclotron with low energy. The IsoDAR project~\cite{Aberle2013} proposes to place such a device underground in the Kamioka 
mine to search for an oscillation pattern in the KamLAND 13-m diameter detector. This would be a disappearance experiment 
directly testing both the reactor and the gallium anomalies starting from a well known  $\anu$ spectrum. In case of positive 
results it would have the ability to disentangle different oscillation models, potentially involving more than one sterile 
neutrino.

\subsubsection{Pion and Kaon Decay at Rest}

For 20 years the puzzling LSND results carried out at LAMPF was never directly tested. This could reliably be achieved by locating a 
detector upstream to a spallation neutron source beam dump. This kind of facility has the advantage to produce a well-understood 
source of electron and muon neutrinos from $\pi^+$ and $\mu^+$ decays-at-rest. The OscSNS project~\cite{Elnimr2013} 
propose to locate a 800-ton gadolinium-doped scintillator detector 60 m away from the Spallation Neutron Source (SNS) at the 
Oak Ridge National Laboratory. The main channel would be the search of the appearance of $\anu$, taking advantage of the 
low duty factor of SNS to reduce cosmic induced backgrounds.

\begin{figure}[h!]
\centering 
\includegraphics[width=1\linewidth]{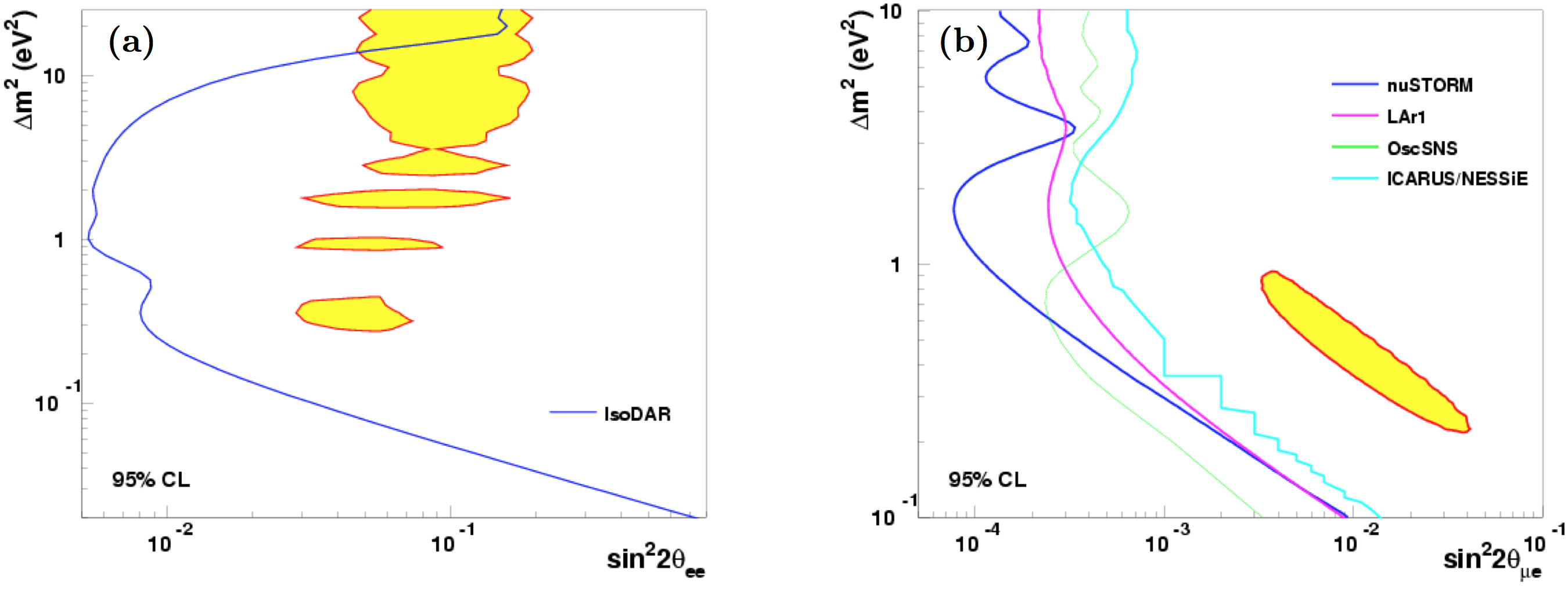}
\caption{\label{fig:2} Projected sensitivity curves for proposed accelerator-based experiments, 
plotted against the global fits~\cite{SnowMass2013}.}
\end{figure}

\subsubsection{Pion Decay in Flight}

To reliably test the LSND and Miniboone anomalies using pion decay in flight neutrinos one would need a setup with two identical
detectors located at different baselines. 
The ICARUS/NESSiE project~\cite{Antonello2012} propose to relocate the 600 tons ICARUS LAr TPC at 1600 m away from 
 a new neutrino beam line at CERN, extracted from the SPS. The near detector would consist of new LAr TPC of 150 tons, 
to be built at 300 m away from the target. This experiment could address both $\nu_e$ and $\anu$ appearance and disappearance channels. It could be complemented by  two muon spectrometers behind each TPC to enhance the ability 
to constrain $\nu_\mu$ dissaperance, a signal expected if the LSND/MiniBooNE anomalies are due to oscillation into sterile neutrino. 
A similar multi-baseline project is being proposed at Fermilab in the Booster Neutrino Beam line. 
The LAr1 project~\cite{Fleming2013} aims to build two LAr TPC, a 40 ton and a 1 kton detector at 100 m and a 1 km for the 
target, respectively. We note that those two projects could be merged in a single experiment, by installing the ICARUS T600 
LAr TPC at Fermilab~\cite{Antonello2013bis}.

\subsubsection{Low Energy Neutrino Factory}

Ultimately precision physics of sterile neutrinos could be done by using a clean and well-understood beam
of $\nu_e$ and $\anumu$ produced by the decay of muons stored into a long storage ring with two 
straight arms pointing to two similar detectors. Furthermore this facility could prototype a future neutrino factory.
As a matter of fact such a neutrino beam could probe precisely both appearance and disappearance processes, 
the golden channel being the search for $\nu_\mu$ appearance from a muon free electron neutrino beam, that is
impossible with in meson decay-in-flight beams. 
The nuSTORM project, based on existing technologies, has been proposed both at CERN~\cite{Adey2013} and 
Fermilab~\cite{Kyberd2013}. Two magnetized iron detectors could be deployed at two different baselines to study 
the golden channel without polluted by wrong sign muons from the beam. 

\section{Summary and Conclusion}
\label{sec:conclusion}
The significance of each short baseline oscillation anomaly is moderate, 
but the concordance of their possible explanation with non-standard 
neutrino oscillation cannot be neglected and calls for new data. 
The projected sensitivity of the experimental proposal discussed in this review are shown in Fig.~\ref{fig:1} for 
the reactor and neutrino generator proposals, and in Fig.~\ref{fig:2} for the accelerator based projects. Data used for these plots have been compiled by the authors of~\cite{SnowMass2013} from the collaborations, as well a from the 
 comprehensive light sterile neutrino  white paper~\cite{WhitePaper2012}.
From these summary plots we see that there is a broad range of sensitivites addressed by the various 
proposals in the appearance and dissapearance oscillation channels. It is likely that reactor and neutrino generator based 
experiments will provide first results since they require less funding and resources. 
The proposed experiments have the potential to test neutrino oscillation transitions with mass-squared difference 
$\Delta m^2>0.1$~eV$^2$ and mixing angle such that $\sin^2 2\theta_{ee} > 0.05$ (see in Fig.~\ref{fig:1}).  
However if sterile neutrino  
oscillations would be confirmed by these first data, then it would be mandatory to study this new physics with a vast 
accelerator-based experimental program, leading to more precise results and accessing  to all possible appearance channels, 
as can bee seen in Fig.~\ref{fig:2}. 
It is worth noting that the observation of neutrino oscillations in at  least two independent detectors employing 
different physics channels, detection methods, and neutrino targets  would be a necessary indication to sign the 
existence of sterile neutrinos. 
First results on the clarification of the short baseline neutrino oscillation anomalies 
might come as early as 2015. 
The situation should be  definitively clarified by 2020, with potentially surprises that 
could lead to major breakthroughs in particle physics, astrophysics, and cosmology.

\section*{Acknowledgements}
Th. Lasserre thanks the European Research Council for support under the Starting Grant StG-307184.

\end{document}